\documentclass[zpreprint,zbstpl]{zeus_paper}

\usepackage[english]{babel}

\chardef\usc=95
\chardef\til=126
\catcode`\@=11 
\DeclareRobustCommand\xdotspace{\futurelet\@let@token\@xdotspace}
\def\@xdotspace{%
  \ifx\@let@token.\else
  \ifx\@let@token\bgroup.\else
  \ifx\@let@token\egroup.\else
  \ifx\@let@token\/.\else
  \ifx\@let@token\ .\else
  \ifx\@let@token~.\else
  \ifx\@let@token!.\else
  \ifx\@let@token,.\else
  \ifx\@let@token:.\else
  \ifx\@let@token;.\else
  \ifx\@let@token?.\else
  \ifx\@let@token/.\else
  \ifx\@let@token'.\else
  \ifx\@let@token).\else
  \ifx\@let@token-.\else
  \ifx\@let@token\@xobeysp.\else
  \ifx\@let@token\space.\else
  \ifx\@let@token\@sptoken.\else
   .\space
   \fi\fi\fi\fi\fi\fi\fi\fi\fi\fi\fi\fi\fi\fi\fi\fi\fi\fi}
\catcode`\@=12 

\newcommand{\stru}[2]{%
   \relax\ifmmode\hbox{\vrule height#1 depth#2 width0pt}%
   \else\vrule height#1 depth#2 width0pt\fi}

\newcommand{\Ronum}[1]{\uppercase\expandafter{\romannumeral#1}}
\newcommand{\ronum}[1]{\expandafter{\romannumeral#1}}
\DeclareRobustCommand{\LaTeXZ}{%
  \LaTeX\kern-.05em4\kern-.1em
  {\raisebox{-0.2ex}{$\scriptstyle\text{ZEUS}$}}\xspace}

\DeclareMathAlphabet{\mathbf}{OT1}{cmr}{bx}{sl}
\newcommand{\eVdist}{\kern-0.06667em}

\newcommand{\Gev}{{\text{Ge}\eVdist\text{V\/}}}

\newcommand{\pb}{\,\text{pb}}

\newcommand{\Tesla}{\,\text{T}}

\newcommand{\slashfrac}[2]{%
  \raisebox{0.5ex}{\ensuremath #1}\kern-0.12em/\kern-0.08em
  \raisebox{-.8ex}{\ensuremath #2}}

\newcommand{\sqr}[3]{%
    {\vcenter{\hrule height.#3ex\hbox{\vrule width.#2ex height#1ex
     \kern#1ex\vrule width.#3ex}\hrule height.#2ex}}}

\catcode`\@=11 
\newcommand{\parenbar}{\mathpalette\p@renb@r}
\def\p@renb@r#1#2{\vbox{%
  \ifx#1\scriptscriptstyle \dimen@.7em\dimen@ii.2em\else
  \ifx#1\scriptstyle \dimen@.8em\dimen@ii.25em\else
  \dimen@1em\dimen@ii.4em\fi\fi \offinterlineskip
  \ialign{\hfill##\hfill\cr
    \vbox{\hrule width\dimen@ii}\cr
    \noalign{\vskip-.3ex}%
    \hbox to\dimen@{$\mathchar300\hfil\mathchar301$}\cr
    \noalign{\vskip-.3ex}%
    $#1#2$\cr}}}
\catcode`\@=12 

\newcommand{\IP}{{\rm I$\kern-0.01667em$P}\xspace}

\mathchardef\qsm=63
\mathchardef\pls=43
\mathchardef\mns=512
\mathchardef\plm=518
\mathchardef\eql=61
\mathchardef\smallleft=300
\mathchardef\smallright=301
\mathchardef\les=316
\mathchardef\gre=318
\mathchardef\leq=532
\mathchardef\grq=533

\catcode`\@=11 
\newcounter{pict@width}
\newcounter{pict@height}
\newlength{\pict@scale}
\newcommand{\psfigadd}[4]{%
\setcounter{pict@width}{1*\ratio{#2+\pict@scale/2}{\pict@scale}}
\setcounter{pict@height}{1*\ratio{#3+\pict@scale/2}{\pict@scale}}
\setlength{\unitlength}{\pict@scale}
\hbox to #2{\hspace{-\fill}\begin{picture}(\thepict@width,\thepict@height)
\put(0,0){\psfig{figure=#1,width=#2,height=#3,clip=}}
\SetScale{0.283466457}
\SetWidth{1.763889}
{#4}
\end{picture}}
}
\newcounter{pict@widthfst}
\newcounter{pict@widthscd}
\newcounter{pict@widthtot}
\newcommand{\psfigaddtwo}[7]{%
\setcounter{pict@widthfst}{1*\ratio{#2+\pict@scale/2}{\pict@scale}}
\setcounter{pict@widthscd}{1*\ratio{#2+#4+\pict@scale/2}{\pict@scale}}
\setcounter{pict@widthtot}{1*\ratio{#2+#4+#6+\pict@scale/2}{\pict@scale}}
\setcounter{pict@height}{1*\ratio{#3+\pict@scale/2}{\pict@scale}}
\setlength{\unitlength}{\pict@scale}
\hbox{\hspace{-\fill}\begin{picture}(\thepict@widthtot,\thepict@height)
\put(0,0){\psfig{figure=#1,width=#2,height=#3,clip=}}
\put(\thepict@widthscd,0){\psfig{figure=#5,width=#6,height=#3,clip=}}
\SetScale{0.283466457}
\SetWidth{1.763889}
{#7}
\end{picture}}
}
\newcommand{\psfigror}[4]{%
\setcounter{pict@width}{1*\ratio{#2+\pict@scale/2}{\pict@scale}}
\setcounter{pict@height}{1*\ratio{#3+\pict@scale/2}{\pict@scale}}
\setlength{\unitlength}{\pict@scale}
\hbox{\begin{picture}(\thepict@width,\thepict@height)
\put(0,\thepict@height){\psfig{figure=#1,width=#3,height=#2,clip=,angle=270}}
\SetScale{0.283466457}
\SetWidth{1.763889}
{#4}
\end{picture}}
}
\newcommand{\psfigrol}[4]{%
\setcounter{pict@width}{1*\ratio{#2+\pict@scale/2}{\pict@scale}}
\setcounter{pict@height}{1*\ratio{#3+\pict@scale/2}{\pict@scale}}
\setlength{\unitlength}{\pict@scale}
\hbox{\begin{picture}(\thepict@width,\thepict@height)
\put(0,0){\psfig{figure=#1,width=#3,height=#2,clip=,angle=90}}
\SetScale{0.283466457}
\SetWidth{1.763889}
{#4}
\end{picture}}
}
\catcode`\@=12 
\newlength\listtextwidth

\catcode`\@=11 
\newlength{\@tabfninsert}
\newlength{\@tabfnwidth}
\newcommand{\tabfootnote}[2]{%
  \setlength{\@tabfninsert}{0.8em}
  \setlength{\@tabfnwidth}{\textwidth}
  \addtolength{\@tabfnwidth}{-\@tabfninsert}
  \addtolength{\@tabfnwidth}{-0.4em}
  \noindent\makebox[\@tabfninsert][r]{\footnotesize$^{#1}$\hfil}\hfill%
  \parbox[t]{\@tabfnwidth}{\footnotesize #2\hfill}}
\catcode`\@=12 

\def\as{\alpha_s}
\def\gp{\gamma p}
\def\eprn{ep\rightarrow e \ {\rm jet} \ {\rm X}}
\def\etjet{E_T^{\rm jet}}
\def\etar{-1<\etajet<2.5}
\def\wrn{$142<\wgp<293$ GeV}
\def\Sjics{Scaled jet invariant cross section}
\def\xtdef{\xt\equiv 2\etjet/\wgp}
\def\wgp{W_{\gp}}
\def\sjics{scaled jet invariant cross section}
\def\set{d\sigma/d\etjet}
\def\xt{x_T}
\def\ejet{E^{\rm jet}}
\def\pp{p\bar p}
\def\oalphass{{\cal O}(\alpha\as)}
\def\asz{\as(\mz)}
\def\etagpr{-2<\etagp<0}
\def\ptmis{p_T^{\rm miss}}
\def\q2{Q^2}
\def\kt{k_T}
\def\etaphi{\eta-\varphi}
\def\etcal{E_{T,{\rm cal}}^{\rm jet}}
\def\etacal{\eta_{\rm cal}^{\rm jet}}
\def\phical{\varphi_{\rm cal}^{\rm jet}}
\def\etacr{-1<\etacal<2.5}
\def\etajet{\eta^{\rm jet}}
\def\phijet{\varphi^{\rm jet}}
\def\asmz#1#2#3#4#5#6{\asz = #1\pm #2\ {\rm (stat.)}\ ^{+#4}_{-#3}\ {\rm (exp.)}\ ^{+#6}_{-#5}\ {\rm (th.)}}
\def\mz{M_Z}
\def\sccs{(\etjet)^4 \ejet d^3\sigma/dp_X^{\rm jet} dp_Y^{\rm jet} dp_Z^{\rm jet}}
\def\etagp{\etajet_{\gp}}
\def\g2{GeV$^2$}
\def\sccsn{(\etjet)^4 \langle \ejet d^3\sigma/dp_X^{\rm jet} dp_Y^{\rm jet} dp_Z^{\rm jet}\rangle_{\eta}}
\def\ele{e^+e^-}
\def\lam#1#2#3#4#5#6{#1\pm #2\ {\rm (stat.)}\ ^{+#4}_{-#3}\ {\rm (exp.)}\ ^{+#6}_{-#5}\ {\rm (th.)}~{\rm GeV}}
\def\bet0#1#2#3#4#5#6{\beta_0 = #1\pm #2\ {\rm (stat.)}\ ^{+#4}_{-#3}\ {\rm (exp.)}\ ^{+#6}_{-#5}\ {\rm (th.)}}
\def\colab#1{#1 Collaboration}
\def\etal{et al.}
\def\bethkenew{S. Bethke, Preprint \mbox{hep-ex/0211012}, 2002}

\begin{document}

\prepnum{{DESY--02--228}}

\title{
Scaling violations and determination of {\boldmath $\as$}\\ from jet
production in {\boldmath $\gp$} interactions at\\ HERA
}                                                       
                    
\author{ZEUS Collaboration}
\date{December 2002}

\abstract{
Differential cross sections for jet photoproduction in the reaction $\eprn$
have been measured with the ZEUS detector at HERA using 82.2~\pb1\ of
integrated luminosity. Inclusive jet cross sections are presented as a
function of the jet transverse energy, $\etjet$, for jets with
$\etjet>17$ GeV and pseudorapidity $\etar$, in the $\gp$
centre-of-mass-energy range \wrn. \Sjics s are presented as a function
of the dimensionless variable $\xtdef$ for $\langle\wgp\rangle=180$
and $255$ GeV. Next-to-leading-order QCD calculations give a good
description of the measured differential cross sections in both
magnitude and shape. The ratio of \sjics s at the two
$\langle\wgp\rangle$ values shows clear non-scaling behaviour. A value
for the strong coupling constant of 
$\asmz{0.1224}{0.0001}{0.0019}{0.0022}{0.0042}{0.0054}$
has been extracted from a QCD analysis of the measured $\set$. The
variation of $\as$ with $\etjet$ is in good agreement with the running
of $\as$ as predicted by QCD.
}

\makezeustitle

\def\3{\ss}

\pagenumbering{Roman}

\begin{center}

{                      \Large  The ZEUS Collaboration              }

\end{center}

  S.~Chekanov,
  D.~Krakauer,
  J.H.~Loizides$^{   1}$,
  S.~Magill,
  B.~Musgrave,
  J.~Repond,
  R.~Yoshida\\
  {\it Argonne National Laboratory, Argonne, Illinois
    60439-4815}~$^{n}$
\par \filbreak

  M.C.K.~Mattingly \\
  {\it Andrews University, Berrien Springs, Michigan 49104-0380}
\par \filbreak

  P.~Antonioli,
  G.~Bari,
  M.~Basile,
  L.~Bellagamba,
  D.~Boscherini,
  A.~Bruni,
  G.~Bruni,
  G.~Cara~Romeo,
  L.~Cifarelli,
  F.~Cindolo,
  A.~Contin,
  M.~Corradi,
  S.~De~Pasquale,
  P.~Giusti,
  G.~Iacobucci,
  A.~Margotti,
  R.~Nania,
  F.~Palmonari,
  A.~Pesci,
  G.~Sartorelli,
  A.~Zichichi  \\
  {\it University and INFN Bologna, Bologna, Italy}~$^{e}$
% ...

\newpage

$^{\    1}$ also affiliated with University College London \\
% ...

\newpage

\begin{tabular}[h]{rp{14cm}}
  
$^{a}$ &  supported by the Natural Sciences and Engineering Research
          Council of Canada (NSERC) \\
$^{b}$ &  supported by the German Federal Ministry for Education and
          Research (BMBF), under contract numbers HZ1GUA 2, HZ1GUB 0,
          HZ1PDA 5, HZ1VFA 5\\
$^{c}$ &  supported by the MINERVA Gesellschaft f\"ur Forschung GmbH,
          the Israel Science Foundation, the U.S.-Israel Binational
          Science Foundation and the Benozyio Center
          for High Energy Physics\\
$^{d}$ &  supported by the German-Israeli Foundation and the Israel
          Science Foundation\\
$^{e}$ &  supported by the Italian National Institute for Nuclear
          Physics (INFN) \\
$^{f}$ &  supported by the Japanese Ministry of Education, Science and
          Culture (the Monbusho) and its grants for Scientific
          Research\\
$^{g}$ &  supported by the Korean Ministry of Education and Korea
          Science and Engineering Foundation\\
$^{h}$ &  supported by the Netherlands Foundation for Research on
          Matter (FOM)\\
$^{i}$ &  supported by the Polish State Committee for Scientific
          Research, grant no. 620/E-77/SPUB-M/DESY/P-03/DZ
          247/2000-2002\\
$^{j}$ &  partially supported by the German Federal Ministry for
          Education and Research (BMBF)\\
$^{k}$ &  supported by the Fund for Fundamental Research of Russian
          Ministry for Science and Edu\-cation and by the German
          Federal Ministry for Education and Research (BMBF)\\
$^{l}$ &  supported by the Spanish Ministry of Education and Science
          through funds provided by CICYT\\
$^{m}$ &  supported by the Particle Physics and Astronomy Research
          Council, UK\\
$^{n}$ &  supported by the US Department of Energy\\
$^{o}$ &  supported by the US National Science Foundation\\
$^{p}$ &  supported by the Polish State Committee for Scientific
          Research, grant no. 112/E-356/SPUB-M/DESY/P-03/DZ
          301/2000-2002, 2 P03B 13922\\
$^{q}$ &  supported by the Polish State Committee for Scientific
          Research, grant no. 115/E-343/SPUB-M/DESY/P-03/DZ
          121/2001-2002, 2 P03B 07022\\
\end{tabular}

\newpage
\pagenumbering{arabic} 
\pagestyle{plain}

\section{Introduction}
Jet production provides a testing ground for the theory of the
strong interaction between quarks and gluons, namely quantum
chromodynamics (QCD). This letter concentrates on one aspect of jet
production, namely, the comparison of jet cross sections for the same
reaction at different centre-of-mass energies. This highlights the
effects of scaling violations, while a QCD analysis of jet-production
rates allows the measurement of the strong coupling constant, $\as$.

The parton model predicts a jet cross section that scales with
the centre-of-mass energy. In this case, the \sjics, $\sccs$, as a
function of the dimensionless variable $\xt\equiv 2\etjet/W$, should be 
independent of $W$, where $W$ is the centre-of-mass energy, $\ejet$ is
the jet energy, $\etjet$ is the jet transverse energy and 
$(p_X^{\rm jet},p_Y^{\rm jet},p_Z^{\rm jet})$ are the components of
the jet momentum. Thus, the ratio of \sjics s for different
centre-of-mass energies will be unity for all $\xt$. On the other
hand, QCD predicts that jet cross sections should exhibit a
non-scaling behaviour, due both to the evolution of the structure
functions of the colliding hadrons and to the running of
$\as$. Scaling violations have been observed in the ratio of the
\sjics s as a function of $\xt$ in $\pp$ collisions at centre-of-mass
energies of either 546 or 630 and 1800~GeV
\cite{prl:70:1376,*prl:86:2523}.

At HERA, similar tests can be made in the photoproduction of jets.
Two types of QCD processes contribute to jet production in $\gamma p$
interactions at $\oalphass$~\cite{pl:b79:83,*np:b166:413,*pr:d21:54,*zfp:c6:241,proc:hera:1987:331,*prl:61:275,*prl:61:682,*pr:d39:169,*zfp:c42:657,*pr:d40:2844}:
either the photon interacts directly with a parton in the proton (the
direct process) or the photon acts as a source of partons, one of
which interacts with a parton in the proton (the resolved
process). Violations of scaling should be observed both in resolved
and direct processes. Furthermore, measurements of high-$\etjet$ jet cross
sections in $\gp$ interactions over a wide range of $\etjet$ allow a
determination of $\asz$ as well as its energy-scale dependence.

This letter presents a measurement of the inclusive jet cross section in
$\gp$ interactions as a function of $\etjet$ in the $\gp$
centre-of-mass-energy range \wrn\ for jets with pseudorapidity
$\etar$. \Sjics s are also presented as a function of $\xt$ for
$\langle\wgp\rangle=180$ and $255$ GeV in the region $\etagpr$, where
$\etagp$ is the jet pseudorapidity in the $\gp$ centre-of-mass frame.

\section{Experimental conditions}
The data were collected during the running period 1998-2000, when 
HERA operated with protons of energy $E_p=920$~GeV and electrons or
positrons of energy $E_e=27.5$~GeV, and correspond to an integrated
luminosity of $82.2\pm 1.9$~\pb1. A detailed description of the ZEUS
detector can be found elsewhere~\cite{pl:b293:465,zeus:1993:bluebook}.
A brief outline of the components that are most relevant for this
analysis is given below.

Charged particles are tracked in the central tracking detector 
(CTD)~\cite{nim:a279:290,*npps:b32:181,*nim:a338:254}, which operates in a
magnetic field of $1.43\Tesla$ provided by a thin superconducting
solenoid. The CTD consists of 72~cylindrical drift-chamber 
layers, organized in nine superlayers covering the
polar-angle\footnote{The ZEUS coordinate system is a right-handed
  Cartesian system, with the $Z$ axis pointing in the proton beam
  direction, referred to as the ``forward direction'', and the $X$
  axis pointing left towards the centre of HERA. The coordinate origin
  is at the nominal interaction point.}
region \mbox{$15^\circ<\theta<164^\circ$}. The transverse-momentum
resolution for full-length tracks can be parameterised as 
$\sigma(p_T)/p_T=0.0058p_T\oplus0.0065\oplus0.0014/p_T$, with $p_T$ in
$\Gev$. The tracking system was used to measure the interaction vertex
with a typical resolution along (transverse to) the beam direction of
0.4~(0.1)~cm and to cross-check the energy scale of the calorimeter.

The high-resolution uranium--scintillator calorimeter
(CAL)~\cite{nim:a309:77,*nim:a309:101,*nim:a321:356,*nim:a336:23} covers 
$99.7\%$ of the total solid angle and consists 
of three parts: the forward (FCAL), the barrel (BCAL) and the rear (RCAL)
calorimeters. Each part is subdivided transversely into towers and
longitudinally into one electromagnetic section (EMC) and either one
(in RCAL) or two (in BCAL and FCAL) hadronic sections (HAC). The
smallest subdivision of the calorimeter is called a cell. Under
test-beam conditions, the CAL single-particle relative energy
resolutions were $\sigma(E)/E=0.18/\sqrt{E\,({\rm GeV})}$ for
electrons and $\sigma(E)/E=0.35/\sqrt{E\,({\rm GeV})}$ for hadrons.

The luminosity was measured from the rate of the bremsstrahlung process 
$e^+p\rightarrow e^+\gamma p$. The resulting small-angle energetic photons
were measured by the luminosity
monitor~\cite{desy-92-066,*zfp:c63:391,*acpp:b32:2025}, a
lead-scintillator calorimeter placed in the HERA tunnel at $Z=-107$ m.

\section{Data selection and jet search}
A three-level trigger system was used to select events
online~\cite{zeus:1993:bluebook,proc:chep:1992:222}. At the first level, events were
triggered by a coincidence of a regional or transverse energy sum in
the CAL and at least one track from the interaction point measured in
the CTD. At the second level, a total transverse energy of at least 
$8$~GeV, excluding the energy in the eight CAL towers immediately
surrounding the forward beampipe, was required, and cuts on CAL
energies and timing were used to suppress events caused by
interactions between the proton beam and residual gas in the
beampipe. At the third level, a jet algorithm was applied to the CAL
cells and jets were reconstructed using the energies and positions of
these cells. Events with at least one jet with $E_T>10$~GeV and
$\eta<2.5$ were accepted.

Events from collisions between quasi-real photons and protons were
selected offline using similar criteria to those reported in a previous 
publication~\cite{pl:b531:9}. The main steps are briefly discussed
here. After requiring a reconstructed event vertex consistent with the
nominal interaction position and cuts based on the tracking
information, the contamination from beam-gas interactions, cosmic-ray
showers and beam-halo muons was negligible. Charged current deep inelastic
scattering (DIS) events were rejected by requiring the total missing
transverse momentum, $\ptmis$, to be small compared to the total
transverse energy, $E^{\rm tot}_T$, i.e.
\mbox{$\ptmis/\sqrt{E^{\rm tot}_T}<2\ \sqrt{\rm GeV}$}.
Any neutral current (NC) DIS events with an identified scattered-positron
or electron candidate in the CAL~\cite{nim:a365:508,*nim:a391:360}
were removed from the sample using the method described
previously~\cite{pl:b322:287}. The remaining background from NC DIS 
events was estimated by Monte Carlo (MC) techniques to be below $0.3\%$
and was neglected. The selected sample consisted of events from $ep$
interactions with $\q2\lesssim 1$ \g2\ and a median
$\q2\approx 10^{-3}$~\g2, where $\q2$ is the virtuality of the
exchanged photon. The events were restricted to $\gamma p$
centre-of-mass energies in the range \mbox{\wrn}, as described in
Section~\ref{secener}.

The longitudinally invariant $\kt$ cluster
algorithm~\cite{np:b406:187} was used in the inclusive
mode~\cite{pr:d48:3160} to reconstruct jets in the hadronic final
state from the energy deposits in the CAL cells. The jet search was
performed in the pseudorapidity-azimuth ($\etaphi$) plane of the
laboratory frame. The jet variables were defined according to the
Snowmass convention~\cite{proc:snowmass:1990:134}. The jets
reconstructed from the CAL cell energies are called calorimetric jets
and the variables associated with them are denoted by $\etcal$,
$\etacal$ and $\phical$. A total of $197\;\! 155$~events with at least
one jet satisfying $\etcal>13$~GeV and $\etacr$ were selected.

\section{Monte Carlo simulation}
\label{monte}
The MC programs PYTHIA~6.1~\cite{cpc:82:74} and HERWIG~5.9~\cite{cpc:67:465}
were used to generate resolved and direct photoproduction events. In
both generators, the partonic processes are simulated using
leading-order matrix elements, with the inclusion of initial- and
final-state parton showers. Fragmentation into hadrons was performed
using the Lund string model~\cite{prep:97:31} as implemented in
JETSET~\cite{cpc:39:347,*cpc:43:367} in the case of PYTHIA, and a cluster
model~\cite{np:b238:492} in the case of HERWIG. The generated events 
were used for calculating energy and acceptance corrections. The
corrections provided by PYTHIA were used as default values and those
given by HERWIG were used to estimate the systematic uncertainties
coming from the treatment of the parton shower and
hadronisation. Samples of PYTHIA including multiparton
interactions~\cite{pr:d36:2019} with a minimum transverse momentum 
for the secondary scatter of 1~GeV~\cite{epj:c2:61} were used to study
the effects of a possible ``underlying event''.

All generated events were passed through the ZEUS detector- and
trigger-simulation programs based on
GEANT~3.13~\cite{tech:cern-dd-ee-84-1}. They were reconstructed and
analysed by the same program chain as the data. The jet search was
performed using the energy measured in the CAL cells in the
same way as for the data. The same jet algorithm was also applied to
the final-state particles; the jets found in this way are referred to
as hadronic jets.

\section{Fixed-order QCD calculations}
\label{nlo}
The QCD calculations, at both leading order (LO) and next-to-leading
order (NLO), used in this analysis are based on the program by Klasen,
Kleinwort and Kramer~\cite{epjdirectcbg:c1:1}. The calculations use the
phase-space-slicing method~\cite{kramer:1984:slicing} with an
invariant-mass cut to isolate the singular regions of the phase
space. The number of flavours was set to five; the renormalisation,
$\mu_R$, and factorisation scales, $\mu_F$, were set to
$\mu_R=\mu_F=\mu=\etjet$; $\as$ was calculated at two loops using
$\Lambda^{(5)}_{\overline{\rm MS}}=220$~MeV, which corresponds to
$\asz=0.1175$. The MRST99~\cite{epj:c4:463,*epj:c14:133}
parameterisations of the parton distribution functions (PDFs) of the
proton and the GRV~\cite{pr:d45:3986,*pr:d46:1973} sets for the photon were used as
defaults for the comparisons with the measured cross sections.

Since the measurements refer to jets of hadrons, whereas the QCD
calculations refer to partons, the predictions were corrected to the
hadron level using the MC models. The multiplicative correction factor,
$C_{\rm had}$, defined as the ratio of the cross section for jets of
hadrons over that for jets of partons, was estimated with the PYTHIA
and HERWIG programs. The values of $C_{\rm had}$ obtained with PYTHIA
were taken as the defaults; the predictions from the two models were in
good agreement. The values of $C_{\rm had}$ differed from unity by less
than $2.5\%$.

\section{Energy and acceptance corrections}
\label{secener}
The comparison of the reconstructed jet variables for the hadronic and 
the calorimetric jets in simulated events showed that no correction
was needed for $\etajet$ and $\phijet$ ($\etajet\simeq\etacal$ and
$\phijet \simeq \phical$). However, the transverse energy of the
calorimetric jet was an underestimate of the corresponding
hadronic jet energy by an average of $\sim 15\%$, with an r.m.s. of
$\sim 10\%$. This underestimation was mainly due to the energy lost by
the particles in the inactive material in front of the CAL. The
transverse-energy corrections to calorimetric jets, as a function of
$\etacal$ and $\etcal$ and averaged over $\phical$, were determined
using the MC events. Henceforth, jet variables without subscript refer
to the corrected values. After these corrections to the jet transverse
energy, events with at least one jet satisfying $\etjet>17$~GeV and
$\etar$ were retained.

The $\gp$ centre-of-mass energy is given by $\wgp=\sqrt{sy}$, where
$y$ is the inelasticity variable and $\sqrt{s}$ is the $ep$ centre-of-mass
energy, $s=4E_eE_p$. The inelasticity variable was reconstructed using the
method of Jacquet-Blondel~\cite{proc:epfacility:1979:391},
$y_{\rm JB}=(E-p_Z)/2E_e$, where $E$ is the total CAL energy and $p_Z$
is the $Z$ component of the energy measured in the CAL cells. The
value of $y$ was systematically underestimated by $\sim 20\%$
with an r.m.s. of $\sim 10\%$. This effect, which was due to energy
lost in the inactive material in front of the CAL and to particles
lost in the rear beampipe, was satisfactorily reproduced by the MC
simulation of the detector. The MC event samples were therefore used
to correct for this underestimation~\cite{epj:c4:591} and obtain
$y_{\rm cor}$. Events were required to have \mbox{\wrn}, where
$\wgp=\sqrt{sy_{\rm cor}}$.

The variable $\xt$ was reconstructed using the corrected values of
$\etjet$ and $\wgp$. Its resolution was $\sim 12\%$. The variable
$\etagp$ was computed by boosting $\etajet$ to the $\gp$
centre-of-mass frame using the formula
$\etagp=\etajet-\ln(2E_p/\wgp)$. The comparison of $\etagp$ for the
hadronic and the calorimetric jets in simulated events showed a good
correlation, so that no correction was needed. The resolution on
$\etagp$ was $\sim 0.08$.

The PYTHIA MC event samples of resolved and direct processes were used to
compute the acceptance corrections to the jet distributions. These
correction factors took into account the efficiency of the
trigger, the selection criteria and the purity and efficiency of the
jet reconstruction. The contributions from direct and resolved
processes in the MC models were added according to a fit to the
uncorrected data distribution of the energy deposited in the RCAL. A
reasonable description of the $\etjet$, $\etajet$, $\wgp$, $\etagp$
and $\xt$ distributions in the data was provided by both PYTHIA and
HERWIG. The differential inclusive jet cross sections were obtained by
applying bin-by-bin corrections to the measured distributions. These
correction factors differed from unity by typically less than $10\%$.

\section{Experimental uncertainties}
\label{expunc}
A detailed study of the experimental systematic uncertainties of the
cross-section measurements included the following sources:

\begin{itemize}
 \item the effect of the presence of a possible underlying event was
   estimated by using the samples of PYTHIA including multiparton
   interactions to evaluate the correction factors. This effect was
   typically $5\%$ and increased to $\sim 10\%$ in the high-$\xt$ tail
   of the \sjics s;
 \item the effect of the treatment of the parton shower and hadronisation
   was estimated by using the HERWIG generator to evaluate the
   correction factors. The uncertainty in the cross sections was
   typically $2\%$;
 \item the effect of the uncertainty on the modelling of the $\q2$
   spectrum of resolved processes in the MC was estimated by using the
   different approximations implemented in PYTHIA and HERWIG. The
   uncertainty in the cross sections was below $2\%$;
 \item the effect of the uncertainty on $\wgp$ was
   estimated by varying $y_{\rm JB}$ by $\pm 1\%$ in simulated events.
   The uncertainty in the cross sections was below $1\%$ at low $\etjet$,
   increasing to $\sim 3\%$ at high $\etjet$;
 \item the effect of the uncertainty on the parameterisations of the
   proton and photon PDFs was estimated by using alternative sets of PDFs 
   in the MC simulation to calculate the correction factors. The variation
   of the cross sections was smaller than $1\%$ in each case.
\end{itemize}

The uncertainty on the simulation of the trigger was negligible. All
the above systematic uncertainties were added in quadrature, giving a
total systematic uncertainty in the cross sections of $5\%$ at low
$\etjet$, increasing to $\sim 10\%$ at high $\etjet$. The absolute
energy scale of the calorimetric jets in simulated events was varied
by its uncertainty of
$\pm 1\%$~\cite{pl:b531:9,epj:c23:615,*hep-ex-0206036}. The effect of
this variation on the inclusive jet cross sections was typically
$\mp 5\%$ at low $\etjet$ increasing to $\mp 10\%$ at high
$\etjet$. This uncertainty is highly correlated between measurements
in different bins. The uncertainty in the luminosity determination of
$2.25\%$ was not included.

\section{Uncertainties on the theoretical predictions}
\label{thunc}
The following uncertainties were considered:
\begin{itemize}
\item the uncertainty on the NLO calculations due to higher-order
  terms was estimated by varying $\mu$ between $\etjet/2$ and
  $2\etjet$. It was less than $10\%$ and mainly affected the
  normalisation. In the ratio of the \sjics s, it was less than $2.5\%$;
\item the uncertainty on the NLO calculations due to the uncertainties on
  the photon PDFs was estimated by using an alternative set of
  parameterisations, AFG-HO~\cite{zfp:c64:621}. The effect was below
  $5\%$ for the cross sections and $2\%$ for the ratio;
\item the uncertainty on the NLO calculations due to the statistical
  and correlated systematic experimental uncertainties of each data set
  used in the determination of the proton PDFs was calculated, making use
  of the results of an analysis~\cite{epj:c14:285} that provided the
  covariance matrix of the fitted PDF parameters and the derivatives
  as a function of Bjorken $x$ and $\mu_F^2$. The resulting
  uncertainty in the cross sections was $1\%$ at low $\etjet$ and
  increased to $5\%$ at high $\etjet$. The uncertainty in the ratio of
  the \sjics s was below $0.3\%$. To estimate the uncertainties on the
  cross sections due to the theoretical uncertainties affecting the
  extraction of the proton PDFs, the calculation of all the
  differential cross sections was repeated using a number of different
  parameterisations obtained under different theoretical assumptions
  in the DGLAP fit~\cite{epj:c14:285}. This uncertainty was below
  $3\%$ for the cross sections and negligible for the ratio;
\item the uncertainty on the NLO calculations due to that on 
  $\asz$ was estimated by varying $\asz$ within its
  uncertainty~\cite{jp:g26:r27} and, simultaneously, by repeating the
  calculations using two additional sets of proton PDFs,
  MRST99$\uparrow\uparrow$ and MRST99$\downarrow\downarrow$,
  determined assuming $\asz=0.1225$ and $0.1125$, respectively. The
  difference between the calculations using these sets and MRST99 was
  scaled by $60\%$ to reflect the current uncertainty on the world
  average of $\as$~\cite{jp:g26:r27}. The resulting uncertainty in the
  cross sections was $\sim 8\%$ at low $\etjet$ decreasing to
  $\sim 2\%$ at high $\etjet$. The uncertainty in the ratio of the 
  \sjics s was below $4\%$;
\item the difference in the hadronisation corrections as predicted by
  PYTHIA and HERWIG resulted in an uncertainty smaller than $2.5\%$.
\end{itemize}

All the above theoretical uncertainties were added in quadrature.

\section{Results}
\label{secres}
\subsection{Inclusive jet differential cross sections}
Using the selected data, inclusive jet differential
cross sections were measured for \wrn. The cross sections were determined
for jets with $\etjet>17$~GeV and $\etar$. There were
$113\;\! 843$~events, containing $145\;\! 797$~jets, in this kinematic
region.

\begin{figure}[p]
\vfill
\setlength{\unitlength}{1.0cm}
\begin{picture} (18.0,18.0)
\put (-2.0,-6.0){\epsfig{figure=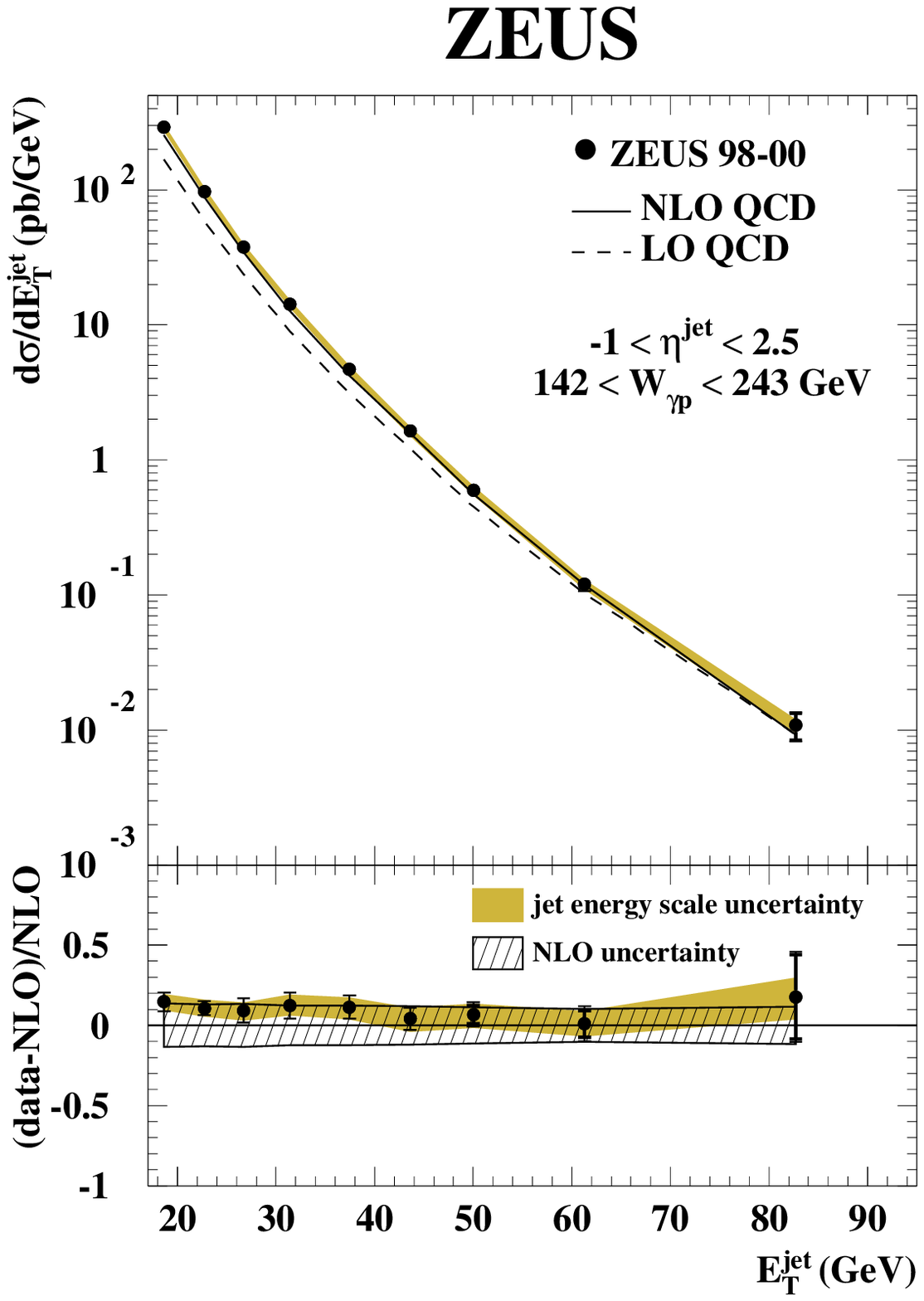,width=21cm}}
\put (12.0,15.0){\bf\small a)}
\put (12.0,5.3){\bf\small b)}
\end{picture}
\vspace{-1.5cm}
\caption
{\it a) Measured inclusive jet cross section, $\set$ (dots).
The thick error bars represent the statistical uncertainties of the
data, and the thin bars show the statistical and systematic
uncertainties added in quadrature. The uncertainty associated with the absolute
energy scale of the jets is shown separately as a shaded band.
The LO (dashed line) and NLO (solid line) QCD parton-level
calculations corrected for hadronisation effects
are also shown. b) The fractional
difference between the measured $\set$ and the NLO QCD calculation;
the hatched band shows the uncertainty of the calculation.}
\label{fig1}
\vfill
\end{figure}

\begin{figure}[p]
\vfill
\setlength{\unitlength}{1.0cm}
\begin{picture} (18.0,18.0)
\put (-2.0,-6.0){\epsfig{figure=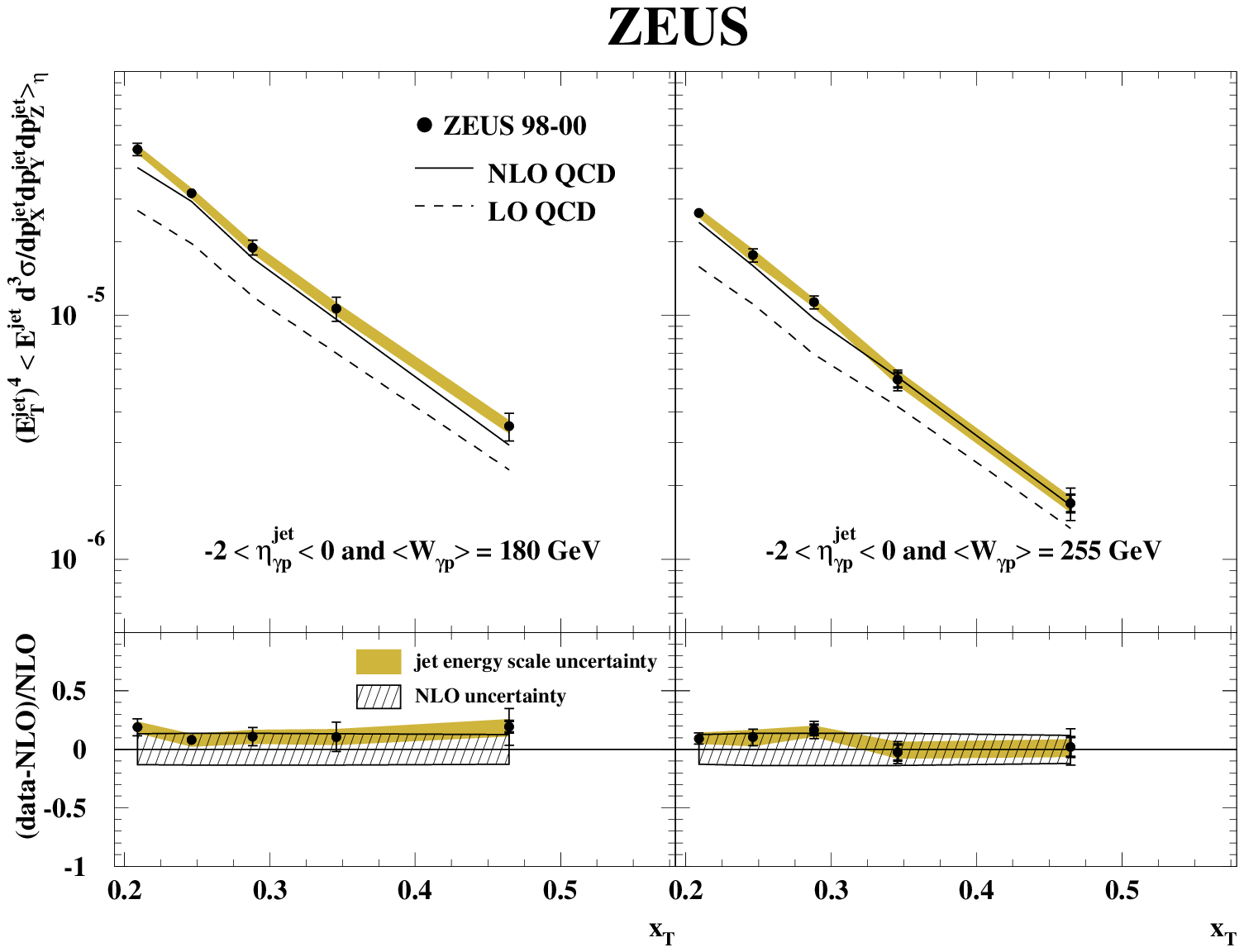,width=21cm}}
\put (7.6,10.9){\bf\small a)}
\put (14.5,10.9){\bf\small b)}
\end{picture}
\caption
{\it Measured \sjics, $\sccsn$, averaged
  over $-2<\etagp<0$, as a function of $\xt$ (dots) for a) 
  $\langle\wgp\rangle=180$ GeV,
  b) $\langle\wgp\rangle=255$ GeV. Other details are as given in the
  caption to Fig.~\ref{fig1}.}
\label{fig2}
\vfill
\end{figure}

The cross-section $\set$, measured in the $\etjet$ range between 17
and 95~GeV, is presented in Fig.~\ref{fig1} and Table~\ref{tabone}.
The data points are located at the weighted mean of each $\etjet$ bin. The
measured $\set$ falls by over five orders of magnitude in this $\etjet$
range. Figure~\ref{fig2} and Table~\ref{tabtwo}
show the scaled jet invariant cross-section, $\sccsn$, averaged over the range $-2<\etagp<0$, 
as a function of $\xt$ for $\langle\wgp\rangle$ values of 180 and 255~GeV;
the $\langle\wgp\rangle$ values were chosen as the
centres of the intervals 169-191~GeV and 240-270~GeV.
The measurements were restricted to the same range in $\etagp$ to have the
same acceptance for the two $\langle\wgp\rangle$ intervals. 

Fixed-order QCD calculations are compared to the data in
Fig.~\ref{fig1}. The LO QCD calculation underestimates the
measured cross section by about $50\%$ for
$\etjet<45$ GeV. The calculation that includes NLO corrections gives a
good description of the data within the experimental and theoretical
uncertainties over the complete $\etjet$ range studied. In particular,
no significant deviation is observed in the highest $\etjet$ region. The
NLO calculations also give a good description of the \sjics s as a
function of $\xt$, as shown in Fig.~\ref{fig2}.

\subsection{Test of scaling}
\begin{figure}[p]
\vfill
\setlength{\unitlength}{1.0cm}
\begin{picture} (18.0,15.0)
\put (-2.0,-7.5){\epsfig{figure=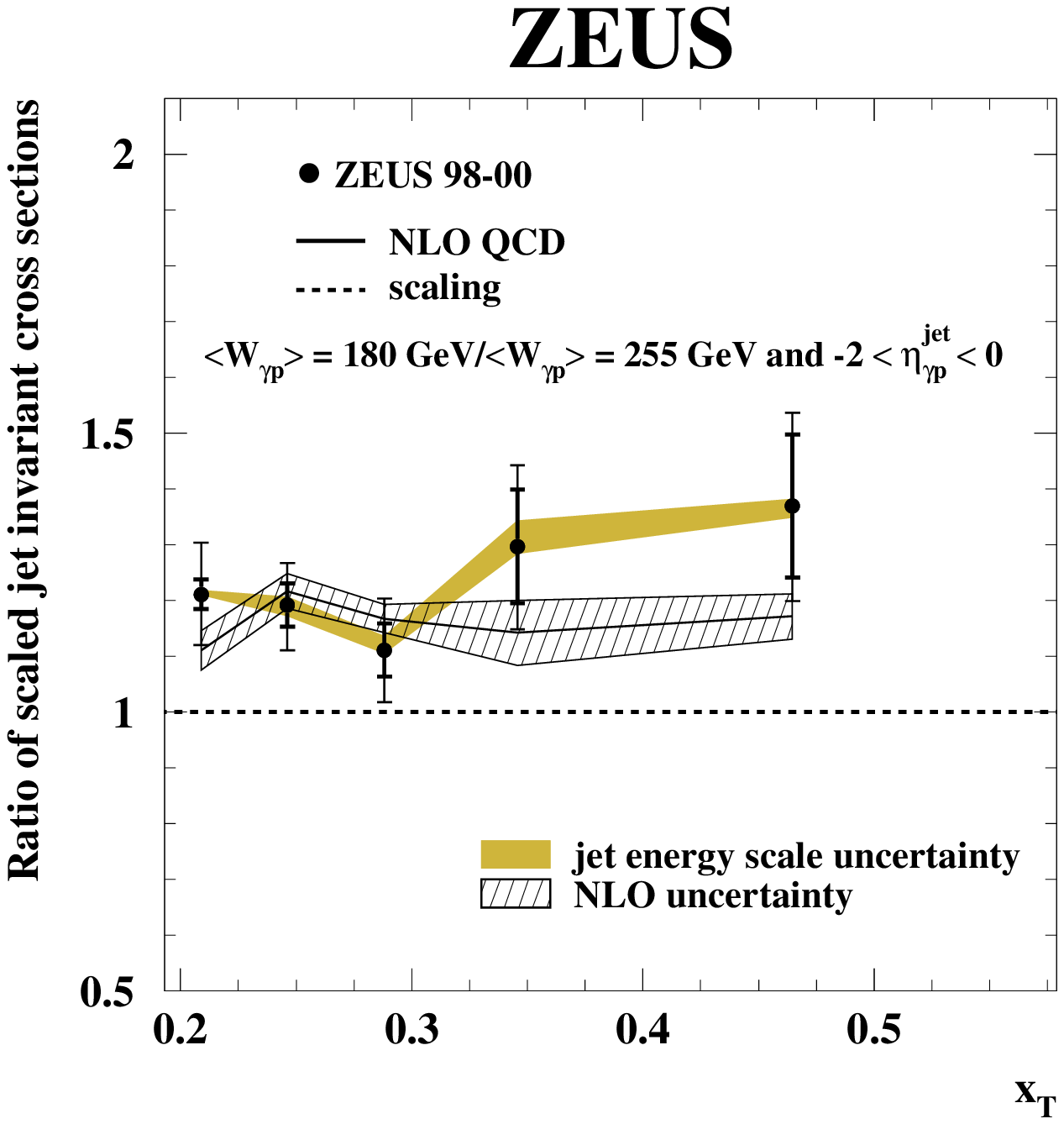,width=21cm}}
\end{picture}
\caption
{\it Measured ratio of \sjics s, after
  correcting for the difference in the photon flux between the two
  $\wgp$ intervals, as a function of $\xt$ (dots). The dashed line is the
  scaling expectation. Other details are as given in the caption to
  Fig.~\ref{fig1}.}
\label{fig3}
\vfill
\end{figure}

To test the scaling hypothesis, the ratio of the \sjics s as a
function of $\xt$ was measured for the two chosen values of
$\langle\wgp\rangle$, after correcting for the difference in the
photon flux~\cite{pl:b319:339} between these intervals. Figure~\ref{fig3} 
and Table~\ref{tabtwo} show the measured ratio as a function of
$\xt$. It shows a clear deviation from unity, in
agreement with the NLO QCD predictions, which include the running of
$\as$ and the evolution of the PDFs with the scale. This constitutes the
first observation of scaling violations in $\gp$ interactions.

The ratio of the \sjics s can be used to test QCD more precisely than
is possible with the individual cross sections, since the
experimental and theoretical uncertainties partially cancel. In
particular, the experimental uncertainty in the absolute energy scale
of the jets cancels almost completely in the ratio. The theoretical
uncertainty on the predictions of the \sjics\ was $13\%$, whereas that
on the ratio was reduced to $2-5\%$. The NLO QCD prediction is in
agreement with the data within the improved experimental (below
$12\%$) and theoretical uncertainties. This agreement shows that the
energy-scale dependence predicted by QCD is in accord with the
measured dependence.

\subsection{Determination of {\boldmath $\asz$}}
\label{aszdet}
The measured cross-section $\set$ as a function of $\etjet$ was used to
determine $\asz$ using the method presented
previously~\cite{pl:b507:70,*pl:b547:164}. The NLO QCD calculations were
performed using the three MRST99 sets of proton PDFs, central, 
MRST99$\downarrow\downarrow$ and MRST99$\uparrow\uparrow$; the value of
$\asz$ used in each partonic cross-section calculation was that associated
with the corresponding set of PDFs. The $\asz$ dependence of the
predicted $\set$ in each bin $i$ of $\etjet$ was parameterised
according to

$$ \left [ \set(\asz) \right ]_i=C_1^i\asz+C_2^i\as^2(\mz),$$
where $C_1^i$ and $C_2^i$ are constants, by using the NLO QCD calculations
corrected for hadronisation effects. Finally, a value of $\asz$ was
determined in each bin of the measured cross section as well as from
all the data points by a $\chi^2$ fit.

The uncertainties on the extracted values of $\asz$ due to the
experimental systematic uncertainties were evaluated by repeating the
analysis for each systematic check presented in
Section~\ref{expunc}. The largest contribution to the experimental
uncertainty comes from the jet energy scale and amounts to $\pm 1.5\%$
on $\asz$. The theoretical uncertainties were evaluated as described in
Section~\ref{thunc}. The largest contribution was the theoretical uncertainty on $\asz$ arising from 
terms beyond NLO, which was ${}^{+4.2}_{-3.3}\%$.
The change of $\asz$ due to the uncertainties on the photon PDFs
and on the hadronisation corrections were $+0.7\%$ and $+0.8\%$,
respectively. The uncertainty on $\asz$ due to the uncertainties on
the proton PDFs was $\pm 0.9\%$. The total theoretical uncertainty on
$\asz$ was obtained by adding these uncertainties in quadrature.

The values of $\as(\mz)$ as determined from the measured $\set$ in
each region of $\etjet$ are shown in Fig.~\ref{fig4}a)
and Table~\ref{tabthree}. By combining all the $\etjet$ regions, the value
of $\asz$ obtained is 
\begin{center}
$\asmz{0.1224}{0.0001}{0.0019}{0.0022}{0.0042}{0.0054}.$
\end{center}
This value of $\asz$ is consistent with the current
world average~\cite{jp:g26:r27} of $0.1183\pm 0.0027$ as well as
with recent determinations from jet production in NC DIS at
HERA~\cite{epj:c19:289,pl:b507:70,*pl:b547:164} and $p\bar{p}$ collisions
at Tevatron~\cite{prl:88:042001}. It has a precision comparable to the
values obtained from $\ele$ interactions~\cite{jp:g26:r27}.

\subsection{Energy-scale dependence of {\boldmath $\as$}}
\begin{figure}[p]
\vfill
\setlength{\unitlength}{1.0cm}
\begin{picture} (18.0,14.5)
\put (-2.0,-7.8){\epsfig{figure=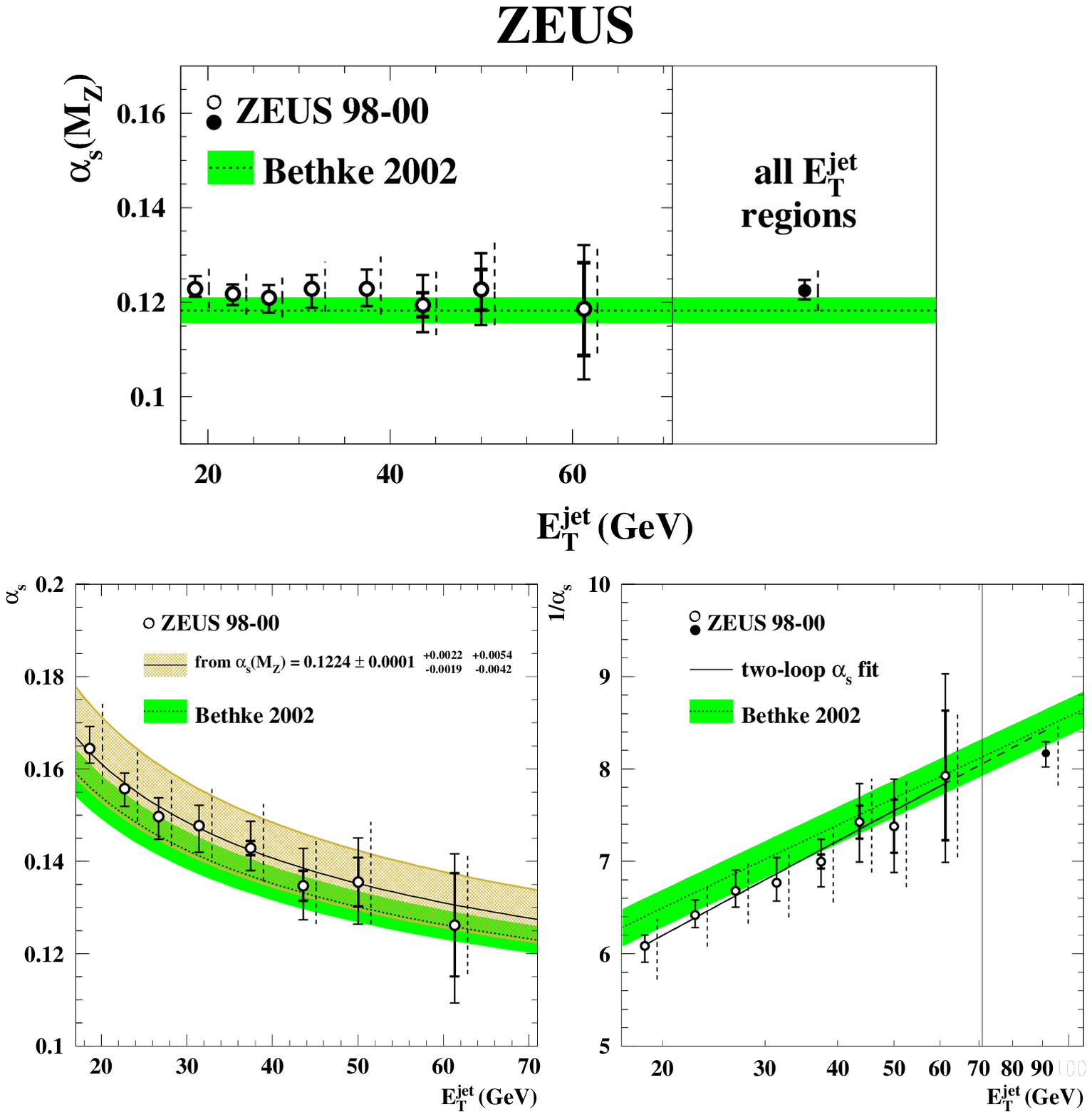,width=21cm}}
\put (12.5,14.0){\bf\small a)}
\put (6.8,6.7){\bf\small b)}
\put (14.6,6.7){\bf\small c)}
\end{picture}
\caption
{\it a) The $\asz$ values determined from the QCD fit of the
  measured $\set$ in the different $\etjet$ regions (open circles). The
  combined value of $\asz$ obtained using all the $\etjet$ regions is
  shown as a dot. b) The $\as(\etjet)$ values
  determined from the QCD fit of the measured $\set$ as a function of
  $\etjet$ (open circles). The solid line represents the prediction of the
  renormalisation group equation obtained from the $\asz$ central
  value as determined in this analysis; the light-shaded area displays
  its uncertainty. c) The $1/\as(\etjet)$ values 
  as a function of $\etjet$ (open circles). The
  solid line represents the result of the two-loop $\as$ fit to the
  measured values. 
  The dashed line represents the extrapolation of the result of
  the fit to $\etjet=\mz$. The dot, plotted at $\etjet=\mz$,
  represents the inverse of the combined value shown in a). In all figures,
  the inner error bars represent the
  statistical uncertainties of the data and the outer error bars show
  the statistical and systematic uncertainties added in
  quadrature. The dashed error bars represent the theoretical
  uncertainties. The current world average~[32] (dotted
  line) and its uncertainty (shaded band) are displayed.}
\label{fig4}
\vfill
\end{figure}

The QCD prediction for the energy-scale dependence of the strong
coupling constant was tested by determining $\as$ from the
measured $\set$ at different $\etjet$ values. The method employed was
the same as described above, but parameterising the $\as$ dependence of
$\set$ in terms of $\as(\langle\etjet\rangle)$ instead of $\asz$,
where $\langle\etjet\rangle$ is the weighted mean of $\etjet$ in each
bin. The measured $\as(\etjet)$ values are shown in Fig.~\ref{fig4}b)
and Table~\ref{tabthree}. The results are in good agreement with the
predicted running of the strong coupling constant over a large range
in $\etjet$.

The energy-scale dependence of the measured $\as(\etjet)$
was quantified by fitting the results using the functional form
predicted by the renormalisation group equation. Perturbative QCD predicts
that $\as^{-1}(\etjet)$ varies approximately linearly with
$\ln\etjet$. At two loops, the energy-scale dependence of
$\as^{-1}(\etjet)$ is given by 
\begin{equation}
 \label{betaqcd}
 \as^{-1}(\etjet) = \frac{\beta_0}{2\pi} \ln(\etjet/\Lambda) \cdot \left [ 
1-\frac{\beta_1}{\beta_0^2}\frac{\ln(2\ln(\etjet/\Lambda))}{\ln(\etjet/\Lambda)}
\right ]^{-1},
\end{equation}
where $\beta_0=11-\frac{2}{3}n_f$, $\beta_1=51-\frac{19}{3}n_f$ and
$n_f$ is the number of active flavours. Thus, the
slope of $\as^{-1}(\etjet)$ gives $\beta_0/2\pi$. A $\chi^2$ fit to the extracted
$\as^{-1}(\etjet)$ values was performed to determine $\beta_0$ using the
functional form given by Eq.~(\ref{betaqcd}), leaving $\beta_0$ and
$\Lambda$ as free parameters; $\beta_1$ was set to $(19\beta_0-107)/2$. 
Figure~\ref{fig4}c) shows the measured $\as^{-1}(\etjet)$ as a
function of $\ln\etjet$ together with the results of the fit. Although
the value of $\Lambda$, 
$\lam{0.535}{0.073}{0.126}{0.140}{0.233}{0.506},$
is not well constrained in the fit, the value of $\asz$ obtained by
extrapolation from the results of the fit is more precise,
$\asmz{0.1188}{0.0009}{0.0039}{0.0043}{0.0067}{0.0069}.$
This determination of $\asz$ is consistent with that of
Section~\ref{aszdet}, in which the running of $\as$ as predicted by
QCD was assumed. The extracted value of $\beta_0$ is
\begin{center}
$\bet0{8.53}{0.22}{0.53}{0.56}{0.82}{1.34}.$
\end{center}
This value is consistent with the prediction of perturbative QCD for
the relevant number of active flavours in the $\etjet$ region
considered, $\beta_0=7.67$ for $n_f=5$. 

\section{Summary and conclusions}
Measurements of differential cross sections for inclusive jet
photoproduction have been made in $ep$ collisions at a centre-of-mass
energy of $318$~GeV using $82.2$~\pb1\ of data collected with the ZEUS
detector at HERA. The cross sections refer to jets identified with
the longitudinally invariant $\kt$ cluster algorithm in the inclusive
mode and selected with $\etjet > 17$~GeV and $\etar$. The
measurements were made in the kinematic region defined by 
$\q2 \; \leq\; 1$~\g2\ and \wrn. 

The inclusive jet cross section was measured as a function of
$\etjet$ in the range between 17 and 95~GeV. The \sjics s, averaged
over $\etagpr$, were measured as a function of the dimensionless
variable $\xt$ for $\langle\wgp\rangle=180$ and $255$~GeV. The NLO QCD
calculations give a good description of the shape and magnitude of the
measured cross sections. No significant deviation with respect to QCD
was observed up to the highest scale studied. The ratio of
\sjics s at two values of $\langle\wgp\rangle$ represents the first
observation of scaling violations in $\gp$ interactions.

A QCD analysis of the measured $\set$ yields a value of
the strong coupling constant of
\begin{center}
$\asmz{0.1224}{0.0001}{0.0019}{0.0022}{0.0042}{0.0054},$
\end{center}
which is in agreement with the current world average and constitutes
the first determination of $\asz$ from jet production in $\gp$
interactions. The value of $\as$ as a function of $\etjet$ is in good
agreement, over a wide range of $\etjet$, with the running of $\as$ as
predicted by QCD.

\newpage
\noindent {\Large\bf Acknowledgements}
\vspace{0.2cm}

We thank the DESY Directorate for their strong support and encouragement.
The remarkable achievements of the HERA machine group were essential for
the successful completion of this work and are greatly appreciated. We
are grateful for the support of the DESY computing and network services.
The design, construction and installation of the ZEUS detector have been
made possible owing to the ingenuity and effort of many people from DESY
and home institutes who are not listed as authors.
We would like to thank M. Klasen for valuable discussions and help in
running his program for calculating QCD jet cross sections in $\gp$
interactions.

\newpage
\clearpage

\begin{mcbibliography}{10}

\bibitem{prl:70:1376}
\colab{CDF}, F. Abe \etal,
\newblock Phys.\ Rev.\ Lett.{} 70~(1993)~1376\relax
\relax
\bibitem{prl:86:2523}
D\O~Collaboration, B. Abbott \etal,
\newblock Phys.\ Rev.\ Lett.{} 86~(2001)~2523\relax
\relax
\bibitem{pl:b79:83}
C.H. Llewellyn Smith,
\newblock Phys.\ Lett.{} B~79~(1978)~83\relax
\relax
\bibitem{np:b166:413}
I. Kang and C.H. Llewellyn Smith,
\newblock Nucl.\ Phys.{} B~166~(1980)~413\relax
\relax
\bibitem{pr:d21:54}
J.F. Owens,
\newblock Phys.\ Rev.{} D~21~(1980)~54\relax
\relax
\bibitem{zfp:c6:241}
M. Fontannaz, A. Mantrach and D. Schiff,
\newblock Z.\ Phys.{} C~6~(1980)~241\relax
\relax
\bibitem{proc:hera:1987:331}
W.J. Stirling and Z. Kunszt,
\newblock {\em Proc.\ HERA Workshop}, R.D.~Peccei~(ed.), Vol.~2, p.~331.
\newblock DESY, Hamburg, Germany (1987)\relax
\relax
\bibitem{prl:61:275}
M. Drees and F. Halzen,
\newblock Phys.\ Rev.\ Lett.{} 61~(1988)~275\relax
\relax
\bibitem{prl:61:682}
M. Drees and R.M. Godbole,
\newblock Phys.\ Rev.\ Lett.{} 61~(1988)~682\relax
\relax
\bibitem{pr:d39:169}
M. Drees and R.M. Godbole,
\newblock Phys.\ Rev.{} D~39~(1989)~169\relax
\relax
\bibitem{zfp:c42:657}
H. Baer, J. Ohnemus and J.F. Owens,
\newblock Z.\ Phys.{} C~42~(1989)~657\relax
\relax
\bibitem{pr:d40:2844}
H.~Baer, J.~Ohnemus and J.F.~Owens,
\newblock Phys.\ Rev.{} D~40~(1989)~2844\relax
\relax
\bibitem{pl:b293:465}
ZEUS \coll, M.~Derrick \etal,
\newblock Phys.\ Lett.{} B~293~(1992)~465\relax
\relax
\bibitem{zeus:1993:bluebook}
{\mbox
ZEUS \coll, U.~Holm~(ed.),
\newblock {\em The {ZEUS} Detector}.
\newblock Status Report}\\
\mbox{(unpublished), DESY (1993),
\newblock available on}\\
  \texttt{http://www-zeus.desy.de/bluebook/bluebook.html}\relax
\relax
\bibitem{nim:a279:290}
N.~Harnew \etal,
\newblock Nucl.\ Inst.\ Meth.{} A~279~(1989)~290\relax
\relax
\bibitem{npps:b32:181}
B.~Foster \etal,
\newblock Nucl.\ Phys.\ Proc.\ Suppl.{} B~32~(1993)~181\relax
\relax
\bibitem{nim:a338:254}
B.~Foster \etal,
\newblock Nucl.\ Inst.\ Meth.{} A~338~(1994)~254\relax
\relax
\bibitem{nim:a309:77}
M.~Derrick \etal,
\newblock Nucl.\ Inst.\ Meth.{} A~309~(1991)~77\relax
\relax
\bibitem{nim:a309:101}
A.~Andresen \etal,
\newblock Nucl.\ Inst.\ Meth.{} A~309~(1991)~101\relax
\relax
\bibitem{nim:a321:356}
A.~Caldwell \etal,
\newblock Nucl.\ Inst.\ Meth.{} A~321~(1992)~356\relax
\relax
\bibitem{nim:a336:23}
A.~Bernstein \etal,
\newblock Nucl.\ Inst.\ Meth.{} A~336~(1993)~23\relax
\relax
\bibitem{desy-92-066}
J.~Andruszk\'ow \etal,
\newblock Preprint \mbox{DESY-92-066}, DESY, 1992\relax
\relax
\bibitem{zfp:c63:391}
ZEUS \coll, M.~Derrick \etal,
\newblock Z.\ Phys.{} C~63~(1994)~391\relax
\relax
\bibitem{acpp:b32:2025}
J.~Andruszk\'ow \etal,
\newblock Acta Phys.\ Pol.{} B~32~(2001)~2025\relax
\relax
\bibitem{proc:chep:1992:222}
W.~H.~Smith, K.~Tokushuku and L.~W.~Wiggers,
\newblock {\em Proc.\ Computing in High-Energy Physics (CHEP), Annecy, France,
  Sept.~1992}, C.~Verkerk and W.~Wojcik~(eds.), p.~222.
\newblock CERN, Geneva, Switzerland (1992).
\newblock Also in preprint \mbox{DESY 92-150B}\relax
\relax
\bibitem{pl:b531:9}
\colab{ZEUS}, S. Chekanov \etal,
\newblock Phys.\ Lett.{} B~531~(2002)~9\relax
\relax
\bibitem{nim:a365:508}
H.~Abramowicz, A.~Caldwell and R.~Sinkus,
\newblock Nucl.\ Inst.\ Meth.{} A~365~(1995)~508\relax
\relax
\bibitem{nim:a391:360}
R.~Sinkus and T.~Voss,
\newblock Nucl.\ Inst.\ Meth.{} A~391~(1997)~360\relax
\relax
\bibitem{pl:b322:287}
ZEUS \coll, M.~Derrick \etal,
\newblock Phys.\ Lett.{} B~322~(1994)~287\relax
\relax
\bibitem{np:b406:187}
S. Catani \etal,
\newblock Nucl.\ Phys.{} B~406~(1993)~187\relax
\relax
\bibitem{pr:d48:3160}
S.D.~Ellis and D.E.~Soper,
\newblock Phys.\ Rev.{} D~48~(1993)~3160\relax
\relax
\bibitem{proc:snowmass:1990:134}
J.E.~Huth \etal,
\newblock {\em Research Directions for the Decade. Proceedings of Summer Study
  on High Energy Physics, 1990}, E.L.~Berger~(ed.), p.~134.
\newblock World Scientific (1992).
\newblock Also in preprint \mbox{FERMILAB-CONF-90-249-E}\relax
\relax
\bibitem{cpc:82:74}
T.~Sj\"ostrand,
\newblock Comp.\ Phys.\ Comm.{} 82~(1994)~74\relax
\relax
\bibitem{cpc:67:465}
G.~Marchesini \etal,
\newblock Comp.\ Phys.\ Comm.{} 67~(1992)~465\relax
\relax
\bibitem{prep:97:31}
B.~Andersson \etal,
\newblock Phys.\ Rep.{} 97~(1983)~31\relax
\relax
\bibitem{cpc:39:347}
T.~Sj\"ostrand,
\newblock Comp.\ Phys.\ Comm.{} 39~(1986)~347\relax
\relax
\bibitem{cpc:43:367}
T.~Sj\"ostrand and M.~Bengtsson,
\newblock Comp.\ Phys.\ Comm.{} 43~(1987)~367\relax
\relax
\bibitem{np:b238:492}
B.~R.~Webber,
\newblock Nucl.\ Phys.{} B~238~(1984)~492\relax
\relax
\bibitem{pr:d36:2019}
T. Sj\"ostrand and M. van Zijl,
\newblock Phys.\ Rev.{} D~36~(1987)~2019\relax
\relax
\bibitem{epj:c2:61}
ZEUS \coll, J.~Breitweg \etal,
\newblock Eur.\ Phys.\ J.{} C~2~(1998)~61\relax
\relax
\bibitem{tech:cern-dd-ee-84-1}
R.~Brun et al.,
\newblock {\em {\sc geant3}},
\newblock Technical Report CERN-DD/EE/84-1, CERN, 1987\relax
\relax
\bibitem{epjdirectcbg:c1:1}
M.~Klasen, T.~Kleinwort and G.~Kramer,
\newblock Eur.\ Phys.\ J.\ Direct{} C~1~(1998)~1\relax
\relax
\bibitem{kramer:1984:slicing}
G.~Kramer,
\newblock {\em Theory of Jets in Electron-Positron Annihilation}.
\newblock Springer, Berlin, (1984)\relax
\relax
\bibitem{epj:c4:463}
A.D.~Martin \etal,
\newblock Eur.\ Phys.\ J.{} C~4~(1998)~463\relax
\relax
\bibitem{epj:c14:133}
A.D.~Martin \etal,
\newblock Eur.\ Phys.\ J.{} C~14~(2000)~133\relax
\relax
\bibitem{pr:d45:3986}
M.~Gl\"uck, E.~Reya and A.~Vogt,
\newblock Phys.\ Rev.{} D~45~(1992)~3986\relax
\relax
\bibitem{pr:d46:1973}
M.~Gl\"uck, E.~Reya and A.~Vogt,
\newblock Phys.\ Rev.{} D~46~(1992)~1973\relax
\relax
\bibitem{proc:epfacility:1979:391}
F.~Jacquet and A.~Blondel,
\newblock {\em Proceedings of the Study for an $ep$ Facility for {Europe}},
  U.~Amaldi~(ed.), p.~391.
\newblock Hamburg, Germany (1979).
\newblock Also in preprint \mbox{DESY 79/48}\relax
\relax
\bibitem{epj:c4:591}
ZEUS \coll, J.~Breitweg \etal,
\newblock Eur.\ Phys.\ J.{} C~4~(1998)~591\relax
\relax
\bibitem{epj:c23:615}
\colab{ZEUS}, S. Chekanov \etal,
\newblock Eur.\ Phys.\ J.{} C~23~(2002)~615\relax
\relax
\bibitem{hep-ex-0206036}
M. Wing (on behalf of the \colab{ZEUS}, in Proceedings for the ``$10th$
  International Conference on Calorimetry in high Energy Physics'',
\newblock Preprint \mbox{hep-ex/0206036}, 2002\relax
\relax
\bibitem{zfp:c64:621}
P.~Aurenche, J.P.~Guillet and M.~Fontannaz,
\newblock Z.\ Phys.{} C~64~(1994)~621\relax
\relax
\bibitem{epj:c14:285}
M.~Botje,
\newblock Eur.\ Phys.\ J.{} C~14~(2000)~285\relax
\relax
\bibitem{jp:g26:r27}
S.~Bethke,
\newblock J.\ Phys.{} G~26~(2000)~R27;\\ updated in \bethkenew\relax
\relax
\bibitem{pl:b319:339}
S. Frixione \etal,
\newblock Phys.\ Lett.{} B~319~(1993)~339\relax
\relax
\bibitem{pl:b507:70}
ZEUS \coll, J.~Breitweg \etal,
\newblock Phys.\ Lett.{} B~507~(2001)~70\relax
\relax
\bibitem{pl:b547:164}
\colab{ZEUS}, S.~Chekanov et al.,
\newblock Phys.\ Lett.{} B~547~(2002)~164\relax
\relax
\bibitem{epj:c19:289}
H1 \coll, C.~Adloff \etal,
\newblock Eur.\ Phys.\ J.{} C~19~(2001)~289\relax
\relax
\bibitem{prl:88:042001}
\colab{CDF}, T. Affolder \etal,
\newblock Phys.\ Rev.\ Lett.{} 88~(2002)~042001\relax
\relax
\end{mcbibliography}
\providecommand{\etal}{et al.\xspace}
\providecommand{\coll}{Collaboration}
\catcode`\@=11
\def\@bibitem#1{%
\ifmc@bstsupport
  \mc@iftail{#1}%
    {;\newline\ignorespaces}%
    {\ifmc@first\else.\fi\orig@bibitem{#1}}
  \mc@firstfalse
\else
  \mc@iftail{#1}%
    {\ignorespaces}%
    {\orig@bibitem{#1}}%
\fi}%
\catcode`\@=12

\newpage
\clearpage
\begin{table}[p]
\begin{center}
\begin{tabular}{|c||c|c|}                \hline
 &  &  \\ 
$\langle\etjet\rangle$ (GeV) & $d\sigma/d\etjet\ \pm\ {\rm stat.}\ \pm\ {\rm syst.}$ (pb) & 
syst. $\etjet$-scale (pb)  \\ 
 &  &  \\ 
\hline\hline
 &  &  \\ 
$18.6$ & $    290 \pm       1 \pm      14$ & $ ( +     11,     -12)$ \\ 
  &  &  \\ 
$22.7$ & $     97 \pm       1 \pm       4$ & $ ( +      4,      -4)$ \\ 
  &  &  \\ 
$26.7$ & $    37.6 \pm      0.3 \pm      2.6$ & $ ( +     1.6,     -1.9)$ \\ 
  &  &  \\ 
$31.4$ & $    14.2 \pm      0.2 \pm      1.0$ & $ ( +     0.8,     -0.7)$ \\ 
  &  &  \\ 
$37.5$ & $     4.7 \pm      0.1 \pm      0.3$ & $ ( +     0.2,     -0.3)$ \\ 
  &  &  \\ 
$43.6$ & $    1.64 \pm     0.06 \pm     0.09$ & $ ( +    0.10,    -0.12)$ \\ 
  &  &  \\ 
$50.0$ & $    0.59 \pm     0.03 \pm     0.03$ & $ ( +    0.03,    -0.04)$ \\ 
  &  &  \\ 
$61.3$ & $   0.120 \pm    0.010 \pm    0.008$ & $ ( +   0.009,   -0.009)$ \\ 
  &  &  \\ 
$82.7$ & $  0.0109 \pm   0.0024 \pm   0.0009$ & $ ( +  0.0011,  -0.0013)$ \\ 
 &  &  \\ 
\hline
\end{tabular}
\caption{
Measured inclusive jet cross-section $\set$.
The statistical and systematic uncertainties $-$not associated with the
absolute energy scale of the jets$-$ are also indicated. The systematic
uncertainties associated to the absolute energy scale of the jets are quoted 
separately. The overall normalization uncertainty of $2.25\%$ is not included.}
\label{tabone}
\end{center}
\end{table}

\newpage
\clearpage
\begin{table}[p]
\begin{center}
\begin{tabular}{|c||c|c|}                \hline
 &  &  \\ 
$\langle\xt\rangle$ & $\sccsn                                   $   $       \pm\ {\rm stat.}\ \pm\ {\rm syst.}$   & 
syst. $\etjet$-scale  \\ 
      &  $(\times 0.389\ 10^6)$ &  $(\times 0.389\ 10^6)$ \\ 
 &  &  \\ 
\hline\hline
\multicolumn{3}{c}{$\langle\wgp\rangle=180$ GeV} \\ \hline \hline
 &  &  \\ 
$0.209$ & $    18.7 \pm      0.2 \pm      1.1$ & $ ( +     0.7,     -0.7)$ \\ 
$0.246$ & $    12.3 \pm      0.2 \pm      0.4$ & $ ( +     0.5,     -0.6)$ \\ 
$0.288$ & $    7.38 \pm     0.15 \pm     0.49$ & $ ( +    0.36,    -0.37)$ \\ 
$0.346$ & $    4.14 \pm     0.15 \pm     0.45$ & $ ( +    0.23,    -0.24)$ \\ 
$0.465$ & $    1.36 \pm     0.06 \pm     0.17$ & $ ( +    0.07,    -0.09)$ \\ 
 &  &  \\ 
\hline\hline
\multicolumn{3}{c}{$\langle\wgp\rangle=255$ GeV} \\ \hline \hline
 &  &  \\ 
$0.209$ & $    10.2 \pm      0.2 \pm      0.4$ & $ ( +     0.4,     -0.4)$ \\ 
$0.246$ & $    6.86 \pm     0.20 \pm     0.40$ & $ ( +    0.36,    -0.44)$ \\ 
$0.288$ & $    4.41 \pm     0.17 \pm     0.22$ & $ ( +    0.12,    -0.20)$ \\ 
$0.346$ & $    2.12 \pm     0.15 \pm     0.14$ & $ ( +    0.19,    -0.10)$ \\ 
$0.465$ & $    0.66 \pm     0.06 \pm     0.08$ & $ ( +    0.04,    -0.05)$ \\ 
 &  &  \\ 
\hline
\multicolumn{3}{c}{ } \\ \hline
 &  &  \\ 
$\langle\xt\rangle$ & Ratio $\pm\ {\rm stat.}\ \pm\ {\rm syst.}$ & 
syst. $\etjet$-scale \\ 
 &  &  \\ 
\hline\hline
 &  &  \\ 
$0.209$ & $    1.21 \pm     0.03 \pm     0.09$ & $ ( +  0.01,  -0.00)$ \\ 
$0.246$ & $    1.19 \pm     0.04 \pm     0.07$ & $ ( +  0.01,  -0.02)$ \\ 
$0.288$ & $    1.11 \pm     0.05 \pm     0.08$ & $ ( +  0.02,  -0.00)$ \\ 
$0.346$ & $    1.30 \pm     0.10 \pm     0.11$ & $ ( +  0.04,  -0.01)$ \\ 
$0.465$ & $    1.37 \pm     0.13 \pm     0.11$ & $ ( +  0.01,  -0.02)$ \\ 
 &  &  \\ 
\hline
\end{tabular}
\caption{
Measured \sjics\ $\sccsn$ as a
  function of $\xt$ for $\langle\wgp\rangle=180$ and $255$ GeV and their
  ratio after correcting for the difference in the photon
  flux. Other details as in the caption to Table~\ref{tabone}.}
\label{tabtwo}
\end{center}
\end{table}

\newpage
\clearpage
\begin{table}[p]
\begin{center}
\begin{tabular}{|c|c||c|c|}                \hline
 &  &  & \\ 
$\etjet$ range & $\asz \pm\ {\rm stat.}\ \pm\ {\rm syst.}\ \pm\ {\rm th.}$ & $\langle\etjet\rangle$ & $\as(\langle\etjet\rangle) \pm\ {\rm stat.}\ \pm\ {\rm syst.}\ \pm\ {\rm th.}$ \\
(GeV) &  & (GeV) &  \\
 &  &  & \\ 
\hline\hline
 &  &  & \\ 
$17-21$ & $0.1229 \pm 0.0001 \ _{-0.0017}^{+0.0026} \ _{-0.0042}^{+0.0055}$ & $18.6$ & $0.1644 \pm 0.0003 \ _{-0.0031}^{+0.0048} \ _{-0.0076}^{+0.0103} $ \\ 
  &  &  & \\ 
$21-25$ & $0.1217 \pm 0.0002 \ _{-0.0023}^{+0.0020} \ _{-0.0043}^{+0.0052}$ & $22.7$ & $0.1558 \pm 0.0004 \ _{-0.0038}^{+0.0033} \ _{-0.0071}^{+0.0087} $ \\ 
  &  &  & \\ 
$25-29$ & $0.1210 \pm 0.0004 \ _{-0.0032}^{+0.0026} \ _{-0.0041}^{+0.0053}$ & $26.7$ & $0.1497 \pm 0.0006 \ _{-0.0049}^{+0.0040} \ _{-0.0064}^{+0.0083} $ \\ 
  &  &  & \\ 
$29-35$ & $0.1228 \pm 0.0006 \ _{-0.0039}^{+0.0030} \ _{-0.0048}^{+0.0058}$ & $31.4$ & $0.1477 \pm 0.0008 \ _{-0.0057}^{+0.0044} \ _{-0.0069}^{+0.0086} $ \\ 
  &  &  & \\ 
$35-41$ & $0.1228 \pm 0.0011 \ _{-0.0034}^{+0.0040} \ _{-0.0053}^{+0.0069}$ & $37.5$ & $0.1429 \pm 0.0015 \ _{-0.0046}^{+0.0056} \ _{-0.0072}^{+0.0095} $ \\ 
  &  &  & \\ 
$41-47$ & $0.1194 \pm 0.0025 \ _{-0.0052}^{+0.0058} \ _{-0.0064}^{+0.0081}$ & $43.6$ & $0.1347 \pm 0.0032 \ _{-0.0066}^{+0.0074} \ _{-0.0082}^{+0.0105} $ \\ 
  &  &  & \\ 
$47-55$ & $0.1227 \pm 0.0043 \ _{-0.0061}^{+0.0064} \ _{-0.0074}^{+0.0105}$ & $50.0$ & $0.1355 \pm 0.0053 \ _{-0.0074}^{+0.0079} \ _{-0.0091}^{+0.0130} $ \\ 
  &  &  & \\ 
$55-71$ & $0.1186 \pm 0.0098 \ _{-0.0111}^{+0.0092} \ _{-0.0093}^{+0.0134}$ & $61.3$ & $0.1262 \pm 0.0112 \ _{-0.0126}^{+0.0106} \ _{-0.0106}^{+0.0153} $ \\ 
 &  &  & \\ 
\hline
\end{tabular}
\caption{
The $\asz$ values determined from the
QCD fit of the measured $\set$ in the different $\etjet$ regions and
the $\as(\langle\etjet\rangle)$ values determined as a function of
$\etjet$. The statistical, systematic and theoretical uncertainties
are also indicated.}
\label{tabthree}
\end{center}
\end{table}

\newpage
\clearpage

\newpage
\clearpage

\newpage
\clearpage

\newpage
\clearpage

\end{document}